# Adaptive Bee Colony in an Artificial Bee Colony for Solving Engineering Design Problems

[1]Tarun Kumar Sharma, [2]Millie Pant, [3]V.P.Singh

[1, 2] Indian Institute of Technology Roorkee, Roorkee, India
[3] Director General, SCET, Saharanpur, India

Email:{taruniitr1; millidma; singhvp3}@gmail.com

**Abstract** – A wide range of engineering design problems have been solved by the algorithms that simulates collective intelligence in swarms of birds or insects. The Artificial Bee Colony or ABC is one of the recent additions to the class of swarm intelligence based algorithms that mimics the foraging behavior of honey bees. ABC consists of three groups of bees namely employed, onlooker and scout bees. In ABC, the food locations represent the potential candidate solution. In the present study an attempt is made to generate the population of food sources (Colony Size) adaptively and the variant is named as A-ABC. A-ABC is further enhanced to improve convergence speed and exploitation capability, by employing the concept of elitism, which guides the bees towards the best food source. This enhanced variant is called E-ABC. The proposed algorithms are validated on a set of standard benchmark problems with varying dimensions taken from literature and on five engineering design problems. The numerical results are compared with the basic ABC and three recent variant of ABC. Numerically and statistically simulated results illustrate that the proposed method is very efficient and competitive.

## 1. Introduction

Evolutionary Programming (EP) [1], Genetic Algorithms (GA) [2], Particle Swarm Optimization (PSO) [3], Differential Evolution (DE) [4], Ant colony optimization (ACO) [5], and the like, are some algorithms to solve many complex real world problems that include engineering, biology, finance, etc [6]-[16]. These algorithms are inspired by the collective intelligent behavior of some species such as school of fishes, colonies of ants, flock of birds and swarm behavior. Researchers modeled this swarm behavior and engineer tested the efficiency and competitiveness of these designed models on complex design problems.

Artificial Bee colony, proposed by Karaboga in 2005 [17]-[18] is the newest algorithm that belongs to the swarm of swarm intelligence algorithms. ABC simulates the foraging behavior of the bee colony and employs this intelligent foraging behavior to solve numerical and engineering design optimization problems.

Karaboga demonstrated the performance of the ABC algorithm [19] is competitive to other population-based algorithm with an advantage of employing fewer control parameters on constrained and unconstrained optimization problem. An important and basic difference between the Artificial Bee Colony and other population based/swarm intelligence algorithms is that in the ABC algorithm the possible solutions represent food sources, not individuals, honeybees where as in other algorithms, like PSO, each possible solution represents an individual of the swarm.

A brief overview of the algorithm is given in section 2.

Like other population based search algorithm, ABC starts with a population of potential candidate solutions (food sources may be flower patch etc.) where the population size (in case of ABC, the number of food sources) are fixed in the beginning of the algorithm. However, practically analyzing it is very rare that the every patch will contain the same number of flowers. Keeping this in mind, in the present study we propose a modified variant of ABC in which the population of food sources change adaptively. The corresponding algorithm is named A-ABC which is further enhanced (E-ABC) by employing the concept of elitism and improving the exploitation capability where the bees are always guided towards the best food source (i.e. the one having the best fitness function value).

The remaining of the paper is organized as follows; in the next section we give a brief overview of the basic ABC algorithm. In section 3, the proposed variants are described. Numerical results are given in section 4 and finally the conclusions are presented in section 5.

## 2. Artificial Bee Colony: An Outline

ABC is one of the newest algorithms based on the foraging behavior of insects. It tries to model natural behavior of real honey bees in food foraging. Honey bees use several mechanisms like waggle dance to optimally locate food sources and to search new ones. Waggle dance is a means of communication among bees by which the successful foragers share the information not only about the direction and distance of the food sources but also about the amount of nectar available to the other foragers. This information exchange among bees helps them in detecting the optimal food locations. In ABC, this collective cooperative behavior of bees is simulated as an optimization algorithm. Since ABC algorithm is simple in concept, easy to implement, and has fewer control parameters, it has been widely used in many fields. ABC algorithm has been applied successfully to a large number of various optimization problems [6], [19], [20]-[33]. The colony of artificial bees contains three groups of bees: employed bees, onlookers and scouts. A bee waiting on the dance area for making a decision to choose a food source is called onlooker and one going to the food source visited by it before is named employed bee. The other kind of bee is scout bee that carries out random search for discovering new sources. The position of a food source represents a possible solution to the optimization problem and the nectar amount of a food source corresponds to the quality (fitness) of the associated solution. In the algorithm, the first half of the colony consists of employed artificial bees and the second half constitutes the onlookers. The number of the employed bees or the onlooker bees is equal to the number of solutions in the population. At the first step, the ABC generates a randomly distributed initial population of *NP* solutions (food source positions), where NP denotes the size of population. Each solution $x_i$ where $i = 1, 2, ..., SN$ is a *D*-dimensional vector, where *D* is the number of optimization parameters. After initialization, the population of the positions (solutions) is subjected to repeated cycles, $C = 1, 2, ..., MCN$ of the search processes of the employed bees, the onlooker bees and scout bees. An employed bee produces a modification on the position (solution) in her memory depending on the local information (visual information) and tests the nectar amount (fitness value) of the new source (new solution). Provided that the nectar amount of the new one is higher than that of the previous one, the bee memorizes the new position and forgets the old one. Otherwise she keeps the position of the previous one in her memory. After all employed bees complete the search process; they share the nectar information of the food sources and their position information with the onlooker bees on the dance area. An onlooker bee evaluates the nectar information taken from all employed bees and chooses a food source with a probability related to its nectar amount. As in the case of the employed bee, she produces a modification on the position in her memory and checks the nectar amount of the candidate source. Providing that its nectar is higher than that of the previous one, the bee memorizes the new position and forgets the old one. An artificial onlooker bee chooses a food source depending on the probability value associated with that food source pi, calculated as Eq. (1):

$$p_i = \frac{fit_i}{\sum_{i=1}^{SN} fit_i} \quad (1)$$

where $fit_i$ is the fitness value of the solution *i* which is proportional to the nectar amount of the food source in the position *i* and *SN* is the number of food sources which is equal to the number of employed bees. In order to produce a candidate food position from the old one in memory, the ABC uses the following Eq. (2):

$$v_{ij} = x_{ij} + \phi_{ij}(x_{ij} - x_{kj}) \quad (2)$$

where $k \in \{1, 2, ..., NP\}$ and $j \in \{1, 2, ..., D\}$ are randomly chosen indexes. Moreover, $k \neq i$. $\emptyset_{ij}$ is a random number between [-1, 1]. It controls the production of neighbor food sources around xij and represents the comparison of two food positions visible to a bee. This can be seen from Eq. (2), as the difference between the parameters of the $x_{ij}$ and $x_{kj}$ decreases, the perturbation on the position $x_{ij}$ decreases, too. Thus, as the search approaches to the optimum solution in the search space, the step length is adaptively reduced. After each candidate source position is produced and evaluated by the artificial bee, its performance is compared with that of its old one. If the new food source has equal or better quality than the old source, the old one is replaced by the new one. Otherwise, the old one is retained. If a position cannot be improved further through a predetermined named "limit", then that food source is assumed to be abandoned. The corresponding employed bee becomes a scout. The abandoned position will be replaced with a new food source found by the scout. Assume that the abandoned source $x_i$, then the scout discovers a new food source to be replaced with $x_i$. This operation can be defined as in Eq. (3):

$$x_i^j = x_{\min}^j + rand()(x_{\max}^j - x_{\min}^j) \quad (3)$$

where $x_{\max}^j$ and $x_{\min}^j$ are upper and lower bounds of parameter *j*, respectively.

**Pseudocode of the ABC Algorithm**

1. Initialize the population of solutions $x_{ij}$ (i=1,2,…,*SN*, j=1,2,…,*D*).
2. Evaluate the population.
3. cycle (represented as *G*) = 1
   **repeat**
4. Produce new solutions (food source positions) using equation (2).
5. Apply the greedy selection process between $x_{ij,G}$ and $v_{ij,G}$.
6. Calculate the probability values $p_i$ for the solutions $x_{ij,G}$ using the equation (1).

In order to calculate the fitness values of solutions the following equation is employed:

$$fit_i = \begin{cases} \dfrac{1}{1+f_i} & if \ f_i \geq 0 \\ 1 + abs(f_i) & if \ f_i < 0 \end{cases}$$

Normalize $p_i$ values into [0, 1].

7. Produce the new solutions (new positions) $v_{ij,G}$ for the onlookers from the solutions $x_{ij,G}$, selected depending on $p_i$, and evaluate them.
8. Apply the greedy selection process for the onlookers between $x_{ij,G}$ and $v_{ij,G}$.
9. Determine the abandoned solution (source), if exists, and replace it with a new randomly produced solution $x_i$ for the scout using the equation (3).
10. Memorize the best food source position (solution) achieved so far.
11. cycle = cycle+1.
12. until cycle = Maximum Cycle Number (MCN).

## 3. Proposed ABC-SAC Algorithms & its variants

In the proposed ABC-SAC algorithm an attempt is to generate adaptive Colony Size. First of all we randomly generate some initial population of solutions, which at a later stage keep on changing adaptively. The Food Source position (*SN*) of subsequent generations is taken as the average of the population size attribute from all individuals in the current population as follows:

| | |
|---|---|
| 1 | Initialize the population of solutions $x_{i,j,G}$. Then declare p as variable (Initialize *p*=0) |
| 2 | for (*i*=0; *i*<FoodNumber; *i*++) |
| 3 | int $p = p +$ Foods[*i*][*D*]; |
| 4 | FoodNumber = (int) (*p*/FoodNumber) + 0.5); |
| 5 | (Check population size for even as number of food sources equals the half of the colony size) |
| 6 | Then follow the steps 2 – 12 described above in the Pseudocode in Section 2. |

It can be observed from the above code that the population of food source (*SN*) depend on p and may vary in every generation. The only thing to be kept in mind is that the population size should be even in number as the number of the employed bees or the onlooker bees is equal to the number of solutions in the population. This is a profitable situation as it may reduce the number of function evaluations leading to a faster convergence. The population generated in each iteration, using the above code, is shown in Fig. 1.

ABC-SAC1: The proposed ABC-SAC is further modified by including in it a concept of elitism, due to which the bees are always guided towards the best food source. Here, the food source is generated as follows:

$$v_{i,j} = x_{best,j} + \phi_{ij}(x_{r1,j} - x_{k,j}) \qquad (4)$$

where $x_{best,j}$ indicates the best food location. The remaining notations have the same meaning as defined in the previous section. This particular modification will further aid in getting a faster convergence. The bees always look for the best solution.

ABC-SAC2: In the other variant of the proposed ABC-SAC2 algorithm is further implemented to improve the exploitation of ABC by using the equation given below:

$$v_{i,j} = x_{i,j} + \phi_{ij}(x_{i,j} - x_{k,j}) + C(x_{best,j} - x_{r1,j}) \qquad (5)$$

where C is taken as 1.5.

This variant is inspired from the Particle Swarm Optimization (PSO) [34] (though the equation is not exactly same) which takes care of global as well as local exploration so that the domain of the problem is thoroughly is explored.

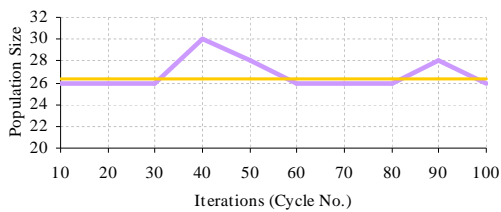

**Figure 1.** Population Graph

## 4. Experimental Settings and Comparison criterion

The ABC has certain control parameters that have to be set by the user. First of all it's the population (number) of food sources (colony size) which is taken as 100 for ABC and for ABC-SAC and its variants ABC-SAC1 and ABC-SAC2. The percentage of onlooker bees is 50% of the colony, the employed bees are 50% of the colony and the number of scout bees is selected as one for ABC. The limit (or generation) is taken 100. Random numbers are generated using inbuilt function *rand*() in DEVC++.

Besides comparing ABC-SAC and its variants ABC-SAC1 and ABC-SAC2 with basic ABC, we have also compared it with a latest variant of ABC called gbest ABC [34] in terms of best and mean fitness function values, standard deviation (SD) and average number of function evaluations (NFE). It should be noted here that according to literature in most of the ABC algorithms the stopping criteria is that of the number of cycles (or generations) however, in the present study we have considered NFE as a stopping criteria. The tests are repeated for 30 runs and the average number of function evaluations (NFE) is recorded. The maximum number of function evaluations is set $10^6$. In every case, a run was terminated when an accuracy of $|f_{max} - f_{min}| < 10^{-20}$ was reached or when the maximum number of function evaluation was reached. Values lesser than $10^{-20}$ are treated as Zero (0). All algorithms are executed on Pentium IV, using DEV C++.

## 5. Result analyses and discussions

The proposed ABC-SAC and its variants are validated on a set of 5 standard, scalable benchmark problems for different dimensions and 5 engineering design problems The test problems and engineering problems are given in the Appendix A and B respectively.

### 5.1. Benchmark Problems

The first four benchmark problems (Sphere, Griewank, Rastringin and Ackley) are tested for dimensions 30 and 60, while the remaining function (Schaffer) is tested for dimensions 2 and 3. The corresponding results are recorded in Tables 1 to 5. From these Tables we can see that all the proposed variants performed quite well in terms of NFE and fitness value. Among the proposed versions, ABC-SAC1 emerged as a clear winner outperforming the other algorithms and the remaining variants significantly. The superior performance of ABC-SAC1 can be further validated with the help of box-plot given in figure 2. The performance curves of selected benchmark problems all the algorithms are illustrated in figure 3. These figures clearly indicate the faster convergence of ABC-SAC. The faster convergence of ABC-SAC1 is due to the presence of elitism which always guides the bees towards the best food source. This increases the convergence rate and results in smaller NFE.

## 5.2. Engineering design problems

Numerical results based on these performance measures are given in Table 6 and Table 7. From Table 6, which gives the average fitness function value and standard deviation, we see that in terms of average fitness function value and standard deviation all the algorithms gave more or less similar results although in some cases the proposed algorithms gave a marginally better performance than basic ABC and ABC-SAC. However if we compare the NFE in Table 7 the superior performance of the proposed algorithms become more evident.

Acceleration rate (AR) [35] is used to compare the convergence speeds between ABC-SAC2 and other algorithms. It is defined as follows:

$$AR = \frac{NFE_{one\,algorithm} - NFE_{other\,algorithm}}{NFE_{one\,algorithm}}\%$$

From Table 7 we can see that the proposed ABC-SAC2 gives the better results for every problem except F3 in the comparison to the other algorithms. Further from the Table 7 it is clearly analysed that the proposed ABC-SAC2 is faster than ABC by 27.82%, faster than ABC-SAC by 24.36% and faster than ABC-SAC1 by 8.24%.

**Table 1.** Best, Mean, Standard Deviation (Std) and Number of Functions Evaluations (NFE) For Sphere Function.

| Algorithm | Best | Mean | SD | NFE |
|---|---|---|---|---|
| D = 30 | | | | |
| ABC | 1.01E-13 | 1.11E-13 | 1.22E-13 | 97140 |
| ABC - SAC | 1.24E-14 | 2.01E-14 | 1.82E-14 | 83720 |
| ABC – SAC1 | 2.61E-15 | 4.44E-15 | 3.57E-15 | 76900 |
| ABC - SAC.2 | 9.02E-15 | 1.88E-14 | 2.23E-14 | 86620 |
| gBest (C =1.5) | 2.02E-14 | 3.22E-14 | 3.25E-15 | 91404 |
| D = 60 | | | | |
| ABC | 8.64E-13 | 9.18E-13 | 4.36E-13 | 109090 |
| ABC - SAC | 5.28E-14 | 7.45E-14 | 1.94E-14 | 91834 |
| ABC – SAC1 | 8.36E-15 | 9.39E-15 | 4.84E-16 | 83195 |
| ABC - SAC.2 | 7.77E-14 | 9.11E-14 | 7.64E-14 | 88039 |
| gBest (C =1.5) | 2.47E-14 | 1.00E-13 | 6.09E-15 | 95301 |

**Table 2.** Best, Mean, Standard Deviation (Std) and Number of Functions Evaluations (NFE) For Griekwank Function.

| Algorithm | Best | Mean | SD | NFE |
|---|---|---|---|---|
| D = 30 | | | | |
| ABC | 3.31E-16 | 4.82E-16 | 8.67E-17 | 47780 |
| ABC - SAC | 3.82E-18 | 5.07E-18 | 9.19E-18 | 40848 |
| ABC – SAC1 | 4.83E-19 | 5.04E-19 | 8.31E-19 | 37905 |
| ABC - SAC.2 | 5.62E-18 | 6.40E-18 | 9.73E-18 | 40131 |
| gBest (C =1.5) | 1.01E-17 | 2.96E-17 | 4.99E-17 | 42831 |
| D = 60 | | | | |
| ABC | 7.59E-15 | 9.54E-15 | 7.15E-16 | 57975 |
| ABC - SAC | 8.65E-18 | 9.30E-18 | 4.80E-18 | 51695 |
| ABC – SAC1 | 7.39E-19 | 8.40E-19 | 1.24E-19 | 43658 |
| ABC - SAC.2 | 6.35E-18 | 9.00E-18 | 9.62E-18 | 40252 |
| gBest (C =1.5) | 6.97E-16 | 7.55E-16 | 4.13E-16 | 51947 |

**Table 3.** Best, Mean, Standard Deviation (Std) and Number of Functions Evaluations (NFE) For Rastringin Function.

| Algorithm | Best | Mean | SD | NFE |
|---|---|---|---|---|
| D = 30 | | | | |
| ABC | 1.38E-13 | 2.19E-13 | 1.81E-13 | 74867 |
| ABC - SAC | 7.98E-15 | 7.98E-15 | 1.69E-15 | 68002 |
| ABC – SAC1 | 5.39E-16 | 6.79E-16 | 1.40E-17 | 61040 |
| ABC - SAC.2 | 6.68E-15 | 8.87E-15 | 9.13E-16 | 65200 |
| gBest (C =1.5) | 1.08E-14 | 1.33E-14 | 2.45E-14 | 71280 |
| D = 60 | | | | |

| Algorithm | Best | Mean | SD | NFE |
|---|---|---|---|---|
| ABC | 2.29E-13 | 6.31E-13 | 2.63E-13 | 99342 |
| ABC - SAC | 1.16E-15 | 4.20E-15 | 2.22E-16 | 73912 |
| ABC – SAC1 | 1.56E-16 | 5.01E-16 | 2.74E-16 | 67402 |
| ABC - SAC.2 | 6.45E-16 | 8.43E-16 | 9.52E-17 | 71035 |
| gBest (C =1.5) | 1.40E-13 | 3.52E-13 | 1.24E-13 | 81430 |

**Table 4.** Best, Mean, Standard Deviation (Std) and Number of Functions Evaluations (NFE) For Ackley Function.

| Algorithm | Best | Mean | SD | NFE |
|---|---|---|---|---|
| D = 30 | | | | |
| ABC | 1.01E-13 | 1.11E-13 | 1.22E-13 | 97140 |
| ABC - SAC | 1.24E-14 | 2.01E-14 | 1.82E-14 | 83720 |
| ABC – SAC1 | 2.61E-15 | 4.44E-15 | 3.57E-15 | 76900 |
| ABC - SAC.2 | 9.02E-15 | 1.88E-14 | 2.23E-14 | 86620 |
| gBest (C =1.5) | 2.02E-14 | 3.22E-14 | 3.25E-15 | 91404 |
| D = 60 | | | | |
| ABC | 8.64E-13 | 9.18E-13 | 4.36E-13 | 109090 |
| ABC - SAC | 5.28E-14 | 7.45E-14 | 1.94E-14 | 91834 |
| ABC – SAC1 | 8.36E-15 | 9.39E-15 | 4.84E-16 | 83195 |
| ABC - SAC.2 | 7.77E-14 | 9.11E-14 | 7.64E-14 | 88039 |
| gBest (C =1.5) | 2.47E-14 | 1.00E-13 | 6.09E-15 | 95301 |

**Table 5.** Best, Mean, Standard Deviation (Std) and Number of Functions Evaluations (NFE) For Schaffer Function.

| Algorithm | Best | Mean | SD | NFE |
|---|---|---|---|---|
| D = 2 | | | | |
| ABC | 0 | 0 | 0 | 48760 |
| ABC - SAC | 0 | 0 | 0 | 41820 |
| ABC – SAC1 | 0 | 0 | 0 | 32992 |
| ABC - SAC.2 | 0 | 0 | 0 | 33080 |
| gBest (C =1.5) | 0 | 0 | 0 | 44500 |
| D = 3 | | | | |
| ABC | 2.18E-16 | 5.76E-16 | 1.62E-17 | 59900 |
| ABC - SAC | 7.78E-18 | 9.48E-18 | 6.30E-18 | 47450 |
| ABC – SAC1 | 2.03E-19 | 7.56E-19 | 2.39E-19 | 39530 |
| ABC - SAC.2 | 2.03E-06 | 6.51E-18 | 2.86E-19 | 35720 |
| gBest (C =1.5) | 1.01E-19 | 1.85E-18 | 1.01E-17 | 52502 |

**Table 6.** Mean of Fitness Function Value and Standard Deviation (Std) for All Algorithms (F-Function)

| F | D | ABC | ABC-SAC | ABC-SAC1 | ABC-SAC2 |
|---|---|---|---|---|---|
| $F_1$ | 2 | 169.844 (0.00021) | 169.849 (0.0080) | 169.846 (1.70e-016) | 169.842 (1.63.e-016) |
| $F_2$ | 3 | 4.2142 (5.08e-007) | 4.2131 (3.19e-016) | 4.20916 (1.42e-017) | 4.20079 (.000101) |
| $F_3$ | 4 | 1.7638e-008 (3.51e-008) | 4.74e-008 (1.87e-018) | 1.8536e-009 8.08e-023) | 2.9014e-009 (2.94e-024) |
| $F_4$ | 10 | -26.0421 (0.446) | -26.01 (0.719) | -26.0317 (0.0398) | -26.0418 (0.6518) |
| $F_5$ | 6 | 2.96e+06 (0.2648) | 2.99e+06 (0.221) | 2.99e+06 (0.0) | 2.791e+06 (0.00319) |

**Table 7.** Comparison of ABC-SAC2 with other Algorithms in terms of NFE's and AR (%), Here 4/1 Implies ABC-SAC2 Vs. ABC, 4/2 Implies ABC-SAC2 Vs. ABC-SAC And 4/3 Implies ABC-SAC2 Vs ABC-SAC1

| F | ABC | ABC-SAC | ABC-SAC1 | ABC-SAC2 | AR 4/1 | AR 4/2 | AR 4/3 |
|---|---|---|---|---|---|---|---|
| $F_1$ | 306 | 289 | 249 | 197 | 35.62 | 31.83 | 20.88 |
| $F_2$ | 838 | 756 | 524 | 513 | 38.78 | 32.14 | 2.10 |
| $F_3$ | 463 | 418 | 317 | 321 | 30.67 | 23.21 | --- |
| $F_4$ | 240000 | 240000 | 240000 | 240000 | 0.00 | 0.00 | 0.00 |
| $F_5$ | 8438 | 8517 | 6913 | 5568 | 34.01 | 34.62 | 19.46 |
| **Average Acceleration Rate (AR(%))** | | | | | 27.82 | 24.36 | 8.24 |

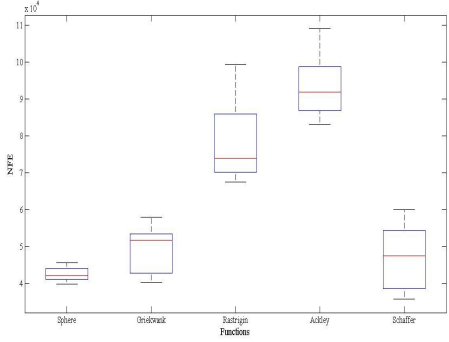

**Figure 2.** Simulation results of all algorithms for benchmark problems

## 6. Conclusions

In the proposed study, we presented the concept of adaptive population of food sources for an Artificial Bee Colony algorithm. The corresponding algorithm named ABC-SAC was further modified by including elitism (ABC-SAC1) and by incorporating global-local exploration (ABC-SAC2) in it. It was observed that by self adapting the food sources, the performance of the basic ABC can be improved. It was also analyzed that this modification can be further improved by considering elitism as the proposed ABC-SAC1 showed better results in comparison to other variants on benchmark problems and even on engineering design problems.

**Appendix A**

| Function | Definition |
|---|---|
| Sphere | $f_1(x) = \sum_{i=1}^{n} x_i^2$ |
|  | S = [-5.12, 5.12], X* = $f_1(0,...,0)$, $f_{min} = 0$ |
| Griekwank | $f_2(x) = \frac{1}{4000}\sum_{i=1}^{n} x_i^2 - \prod_{i=1}^{n} \cos(\frac{x_i}{\sqrt{i}}) + 1$ |
|  | S = [-600, 600], X* = $f_2(0,...,0)$, $f_{min} = 0$ |
| Ackly's | $f_3(x) = -20 * \exp\left(-.2\sqrt{1/n \sum_{i=1}^{n} x_i^2}\right) - \exp\left(1/n \sum_{i=1}^{n} \cos(2\pi x_i)\right) + 20 + e$ |
|  | S = [-32, 32], X* = $f_3(0,...,0)$, $f_{min} = 0$ |
| Restrigin's | $f_4 = 10n + \sum_{i=1}^{n}\left(x_i^2 - 10\cos(2\pi x_i)\right)$ |
|  | S = [-5.12, 5.12], X* = $f_4(0,...,0)$, $f_{min} = 0$ |
| Schaffer | $f_5 = 0.5 + \frac{\sin^2\left(\sqrt{\sum_{i=1}^{n} x_i^2}\right) - 0.5}{\left(1 + 0.001\left(\sum_{i=1}^{n} x_i^2\right)\right)^2}$ |
|  | S = [-100, 100], X* = $f_5(0,...,0)$, $f_{min} = 0$ |

Notes: $f_{min}$ = the minimum of the value of the function: X* = the minimum: S = the feasible region: and $S \in R^n$, n is dimension of the problem

**Appendix B**

Engineering Design Problems

The credibility of an optimization algorithm also depends on its ability to solve real life/engineering's problems. In this paper we took four real life engineering design problems to validate the efficiency of all the proposed algorithms. Mathematical models of problems are given below:

$F_1$: Optimal Capacity of Gas Production Facilities [36]

This is the problem of determining the optimum capacity of production facilities that combine to make an oxygen producing and storing system. Oxygen for basic oxygen furnace is produced at a steady state level. The demand for oxygen is cyclic with a period of one hour, which is too short to allow an adjustment of level of production to the demand. Hence the manager of the plant has two alternatives.

He can keep the production at the maximum demand level; excess production is lost in the atmosphere.

He can keep the production at lower level; excess production is compressed and stored for use during the high demand period.

The mathematical model of this problem is given by:
Minimize:

$$f(x) = 61.8 + 5.72x_1 + 0.2623[(40 - x_1)\ln(\frac{x_2}{200})]^{-0.85}$$
$$+ 0.087(40 - x_1)\ln(\frac{x_2}{200}) + 700.23x_2^{-0.75}$$

Subject to: $x_1 \geq 17.5$, $x_2 \geq 200$; $17.5 \leq x_1 \leq 40$, $300 \leq x_2 \leq 600$.

$F_2$: Optimal Thermohydraulic Performance of an Artificially Roughened Air Heater [37]

In this problem the optimal thermohydraulic performance of an artificially roughened solar air heater is considered. Optimization of the roughness and flow parameters (p/e, e/D, Re) is considered to maximize the heat transfer while keeping the friction losses to be minimum. The mathematical model of this problem is given by:

Maximize $L = 2.51 * \ln e^+ + 5.5 - 0.1 R_M - G_H$

where $R_M = 0.95 x_2^{0.53}$; $G_H = 4.5(e^+)^{0.28}(0.7)^{0.57}$

$e^+ = x_1 x_3 (\bar{f}/2)^{1/2}$; $\bar{f} = (f_s + f_r)/2$

$f_s = 0.079 x_3^{-0.25}$; $f_r = 2(0.95 x_3^{0.53} + 2.5 * \ln(1/2 x_1)^2 - 3.75)^{-2}$

Subject to:

$0.02 \leq x_1 \leq 0.8$, $10 \leq x_2 \leq 40$, $3000 \leq x_3 \leq 20000$

F3: Design of Gear Train [38]

This problem is to optimize the gear ratio for the compound gear train. It is to be designed such that the gear ratio is as close as possible to 1/6.931. For each gear the number of teeth must be between 12 and 60. Since the number of teeth is to be an integer, the variables must be integers. The mathematical model of gear train design is given by,

Minimize $f = \left\{ \dfrac{1}{6.931} - \dfrac{T_d T_b}{T_a T_f} \right\}^2 = \left\{ \dfrac{1}{6.931} - \dfrac{x_1 x_2}{x_3 x_4} \right\}^2$

Subject to: $12 \leq x_i \leq 60$  $i = 1,2,3,4$

$[x_1, x_2, x_3, x_4] = [T_d, T_b, T_a, T_f]$, $x_i$'s should be integers. $T_a$, $T_b$, $T_d$, and $T_f$ are the number of teeth on gears A, B, D and F respectively. The design of a compound gear train is shown in Fig. 1.

F4: The lennard-jones atomic cluster problem [39]

The Lennard-Jones (LJ) model of inert gas cluster has been investigated intensively as a challenging testing ground for global optimization algorithms. The LJ problem can be formulated as follows: Let N atoms be given in three dimensional space. Let $x^i = (x_1^i, x_2^i, x_3^i)^T$ represent the coordinates of atom i. let $X = ((x^1)^T, ..., (x^N)^T)^T$ the LJ potential energy function $v(r_{ij})$ of a pair of atoms (i, j) is given by

$v(r_{ij}) = \dfrac{1}{r_{ij}^{12}} - \dfrac{1}{r_{ij}^6}, 1 \leq i,j \leq N$ where $r_{ij} = ||x^i - x^j||$

The LJ potential function for a single pair of neutral atoms is a simple unimodal function. This is illustrated by Fig. 2. It is easy to find the overall minimum of this function that is assumed at 1 with energy −1. In a complex system, many atoms interact and we need to sum up the LJ potential functions for each pair of atoms in a cluster. The result is a complex energy landscape with many local minima. It is given by:

Minimize $V(X) = \sum_{i<j} v(||x^i - x^j||) = \sum_{i=1}^{N-1} \sum_{j=i+1}^{N} \left( \dfrac{1}{||x^i - x^j||^{12}} - \dfrac{1}{||x^i - x^j||^6} \right)$

As it is known the LJ problem is a highly nonlinear, non convex function with numerous local minimum that takes the LJ problem a challenging standard test problem for the global optimization algorithms. For illustration, we fix one atom's position and let the others move around the fixed one. A cluster of more than 20 atoms has many of local minima along its LJ PES.

F5: Gas transmission compressor design [37]

Minimize $f(x) = 8.61 * 10^5 * x_1^{1/2} x_2 x_3^{-2/3} (x_2^2 - 1)^{-1/2} +$

$3.69 * 10^4 * x_3 + 7.72 * 10^8 * x_1^{-1} x_2^{0.219} - 765.43 * 10^6 * x_1^{-1}$

Bounds: $10 \leq x_1 \leq 55, 1.1 \leq x_2 \leq 2, 10 \leq x_3 \leq 40$

## References

[1] L. J. Fogel, A. J. Owens, M. J. Walsh, Artificial intelligence through a simulation of evolution, edited by M. Maxfield, Callahan, and L. J. Fogel, Biophysics and Cybernetic systems, in: 2nd Cybernetic Sciences Symposium 1965, pp. 131–155.

[2] D. Goldberg, Genetic Algorithms in Search Optimization and Machine Learning, Addison Wesley Publishing Company, Reading, Massachutes 1986.

[3] J. Kennedy, R. C. Eberhart, Particle Swarm Optimization, in: IEEE International Conference on Neural Networks, Perth, Australia, IEEE Service Center, Piscataway, NJ 1995, pp. 1942–1948.

[4] K. Price, R. Storn, Differential Evolution – a Simple and Efficient Adaptive Scheme for Global Optimization Over Continuous Spaces, Technical Report, International Computer Science Institute, Berkley 1995.

[5] M. Dorigo, V. Maniezzo, A. Colorni, Positive feedback as a search strategy, Technical Report 91-016, Politecnico di Milano, Italy, 1991.

[6] Tarun Kumar Sharma, Millie Pant, V.P. Singh, Improved Local Search in Artificial Bee Colony using Golden Section Search, Journal of Engineering, 1:1(2012) 14-19.

[7] Ali Yazdekhasti, Iman Sadeghkhani, Optimal Tuning of TCSC Controller Using Particle Swarm Optimization, Advances in Electrical Engineering Systems, 1:1(2012) 24-29.

[8] Yudong Zhang, Lenan Wu, Rigid Image Registration by PSOSQP Algorithm, Advances in Digital Multimedia, 1:1(2012) 4-8.

[9] Maziar Rezaei Rad, Mani Rezaei Rad, Shahabeddin Akbari, Seyyed Abbas Taher, Using ANFIS, PSO, FCN in Cooperation with Fuzzy Controller for MPPT of Photovoltaic Arrays, Advances in Digital Multimedia, 1:1(2012) 37-45.

[10] Zheng Zhang, Pattern Recognition by PSOSQP and Rule based System, Advances in Electrical Engineering Systems, 1:1(2012) 30-34.

[11] Shuihua Wang, Lenan Wu, An Improved PSO for Bankruptcy Prediction, Advances in Computational Mathematics and its Applications, 1:1(2012) 1-6.

[12] Yudong Zhang, Lenan Wu, Tabu Search Particle Swarm Optimization used in Cluster Analysis, Journal of Science, 1:1(2012) 6-12.

[13] Amir Ghoreishi, Arash Ahmadivand, State Feedback Design Aircraft Landing System with Using Differential Evolution Algorithm, Advances in Computer Science and its Applications, 1:1(2012) 16-20.

[14] Yudong Zhang, Lenan Wu, A Robust Hybrid Restarted Simulated Annealing Particle Swarm Optimization Technique, Advances in Computer Science and its Applications, 1:1(2012) 5-8.

[15] Y. Zhang, L. Wu, Artificial Bee Colony for Two Dimensional Protein Folding, Advances in Electrical Engineering Systems, 1:1(2012) 19-23.

[16] H. Noormohamadi, A.A. Gharaveisi, M. Suresrafil, Novel Bacterial Foraging Algorithm for Optimization Problems, Advances in Electrical Engineering Systems, 1:1(2012) 41-48.

[17] D. Karaboga, An idea based on honey bee swarm for numerical optimization. Technical Report-TR06, Kayseri, Turkey: Erciyes University; 2005.

[18] D. Karaboga, B. Basturk, A powerful and efficient algorithm for numerical function optimization: artificial bee colony (ABC) algorithm, Journal of Global Optimization 39(2007) 171–459.

[19] D. Karaboga, B. Basturk, On the performance of artificial bee colony (ABC) algorithm, Applied Soft Computing, 8(2008) 687-697.

[20] D. Karaboga, C. Ozturk, N. Karaboga, B. Gorkemli, Artificial bee colony programming for symbolic regression, Information Sciences (2012), http://dx.doi.org/10.1016/j.ins.2012.05.002.

[21] M Ma, J. Liang, M. Guo, Y. Fan, Y. Yin, SAR image segmentation based on Artificial Bee Colony algorithm, Applied

Soft Computing, (In Press) doi:10.1016/j.asoc.2011.05.039, in press.

[22] WC. Yeh, TJ. Hsieh, Artificial bee colony algorithm-neural networks for s-system models of biochemical networks approximation. Neural Comput Appl. (2012) doi:10.1007/s00521-010-0435-z.

[23] F. Gao, Feng-xia Fei, Q. Xu, Y. fang Deng, Yi-bo Qi, I. Balasingham, A novel artificial bee colony algorithm with space contraction for unknown parameters identification and time-delays of chaotic systems, Appl. Math. Comput. (2012), http://dx.doi.org/10.1016/j.amc.2012.06.040.

[24] H. Zhang, Y. Zhu, W. Zou, X. Yan, A hybrid multi-objective artificial bee colony algorithm for burdening optimization of copper strip production, Applied Mathematical Modelling, 36:6(2012) 2578-2591.

[25] X. Liao, J. Zhou, R. Zhang, Y. Zhang, An adaptive artificial bee colony algorithm for long-term economic dispatch in cascaded hydropower systems, International Journal of Electrical Power & Energy Systems, 43:1(2012) 1340-1345.

[26] H. Gozde, M. Cengiz Taplamacioglu, İ. Kocaarslan, Comparative performance analysis of Artificial Bee Colony algorithm in automatic generation control for interconnected reheat thermal power system, International Journal of Electrical Power & Energy Systems 42:1(2012) 167-178.

[27] Ali R. Yildiz, A new hybrid artificial bee colony algorithm for robust optimal design and manufacturing, Applied Soft Computing, In Press, 2012.

[28] O. Kisi, C. Ozkan, B. Akay, Modeling discharge–sediment relationship using neural networks with artificial bee colony algorithm, Journal of Hydrology, 428–429(2012) 94-103.

[29] S. K. Mandal, Felix T.S. Chan, M.K. Tiwari, Leak detection of pipeline: An integrated approach of rough set theory and artificial bee colony trained SVM, Expert Systems with Applications, 39:3(2012) 3071-3080.

[30] T.K. Sharma, M. Pant, Enhancing the food locations in an artificial bee colony algorithm, in: IEEE Swarm Intelligence Symposium (SIS), Paris, France, 2011, pp. 119-123.

[31] T.K. Sharma, M. Pant, Enhancing different phases of artificial bee colony for continuous global optimization problems, in: International Conference on Soft Computing for Problem Solving, SocProS 2011, AISC of Advances in Intelligent and Soft Computing, Roorkee, India, Vol. 130, 2011, pp. 715–724.

[32] T.K. Sharma, M. Pant, J.C. Bansal, Artificial Bee Colony with Mean Mutation Operator for Better Exploitation, in: IEEE World Congress on Computational Intelligence (CEC), Brisbane, Australia, 2012, pp. 3050 - 3056.

[33] T.K. Sharma, M. Pant, J.C. Bansal, Some Modifications to Enhance the Performance of Artificial Bee Colony, in: IEEE World Congress on Computational Intelligence (CEC), Brisbane, Australia, 2012, pp. 3454 - 3461.

[34] Zhu G., Kwong S.: Gbest-Guided Artificial Bee Colony Algorithm for Numerical Function Optimization, Appl. Math. Comput. (2010).

[35] S. Rahnamayan, H.R. Tizhoosh, and Salama, M. M. Ali, Opposition-Based Differential Evolution. IEEE Transactions on Evolutionary Computation, 12:1(2008) 64– 79.

[36] [12]C. Beightler, D. Phillips. Applied geometric programming. John Wiley and Sons, New York, 1976.

[37] B. Prasad, J. Saini, Optimal thermo hydraulic performance of artificially roughened solar air heater. Journal Solar Energy, 47(1991) 91–96.

[38] B. Babu, New optimization techniques in engineering. Springer-Verlag, Berlin Heidelberg, 2004.

[39] U. chakraborthy, S. Das, A. konar. Differential evolution with local neighborhood, in: IEEE congress on evolutionary computation, Canada, 2006, 2042–2049.

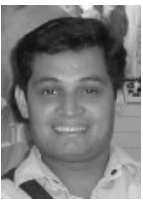

**Tarun Kumar Sharma** did his MCA in 2001, M.Tech (IT) in 2009 and presently pursuing Ph.D from Indian Institute of Technology (IIT) Roorkee, India. He has almost 9 years of teaching experience in Engineering College. His key areas are Evolutionary Computing; Software Engineering; Computer based Optimization Techniques; ERP. His research interest includes swarm intelligence algorithms and their applications in various complex engineering design problems. His publications are in Journals and International Conferences of repute. He volunteered in SocPros-2011, an First International Conference on Soft Computing for Problem solving. He is peer reviewer of many IEEE conferences and International Journals. He is student member of Machine Intelligence Research (MIR) Labs, WA, USA.

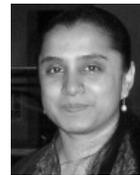

**Millie Pant** is working as an assistant professor in Department of Paper Technology, Indian Institute of Technology (IIT), Roorkee, India since 2007. Her research interest includes evolutionary and swarm intelligence algorithms and their applications in various complex engineering design problems. Her publications are in Journals and International Conferences of repute. She has published over 100 referred on evolutionary algorithms (GA, PSO, DE and ABC) and their applications in electrical design problems, image processing papers. She has been program committee member of over 10 International events and Program Committee Chair of SoCProS-211. She is Program Committee Chair of the 7th International Conference on Bio-Inspired Computing: Theories and Application (BIC-TA 2012) and SoCProS-2012 (International Conference on Soft Computing for Problem Solving).

**V.P. Singh** received the Bachelor's degree from Meerut University, India in the year 1970, Master's and Ph. D degrees in Applied mathematics from the University of Roorkee (now, Indian Institute of Technology Roorkee), India, in 1972, and 1978 respectively. Prof. Singh is currently with SCET Saharanpur, as a Director General. Previously he was Professor in the Department of Paper Technology, IIT Roorkee. His special fields of interests include Applied and Industrial Mathematics, Mathematical Modelling of Pulp washing problems. He has number of publications in journal of repute and has been reviewer for number of Journals and Conferences.